\documentclass[bibnotes,showpacs,twocolumn,aps,prl,a4paper]{revtex4}

\usepackage{color,epsfig,rotating,graphicx,subfigure}
\usepackage{amsmath,amssymb,amscd}

\usepackage[latin1]{inputenc}
\usepackage[english]{babel}

\usepackage{dsfont}
\usepackage{textcomp}

\renewcommand{\Im}{{\rm Im}}

\newcommand{\ri}{{\rm i}}
\newcommand{\rF}{{\rm F}}
\newcommand{\rd}{{\rm d}}
\newcommand{\re}{{\rm e}}

\newcommand{\rp}{{\rm p}}

\newcommand{\kb}{k_{\rm B}}
\newcommand{\rth}{{\rm th}}

\begin{document}

%
%
\title{A mesoscopic description of radiative heat transfer at the nanoscale}

\author{S.-A. Biehs, E. Rousseau, and J.-J. Greffet}

\affiliation{Laboratoire Charles Fabry, Institut d'Optique, CNRS, Universit\'{e} Paris-Sud, Campus
Polytechnique, RD128, 91127 Palaiseau Cedex, France}

\date{09.11.2010}


\begin{abstract}
We present a formulation of the nanoscale radiative heat transfer (RHT) using concepts of mesoscopic physics. 
We introduce the analog of the Sharvin conductance using the quantum of thermal conductance. The formalism provides a convenient framework to analyse the physics of RHT at the nanoscale. Finally, we propose a RHT experiment in the regime of quantized conductance.
\end{abstract}

\maketitle
\newpage

%
%

%
%

It has been discovered in the late sixties that the RHT between two metallic parallel plates can be larger than predicted using the blackbody radiation form~\cite{Cravalho, Hargreaves,Tien}. It is now known that this anomalous RHT is due to the contribution of evanescent waves and becomes significant when the distance separating the interfaces becomes smaller than the thermal wavelength $\lambda_{\rm th}=  \frac{\hbar c}{\kb T}$ where $\hbar$ is Planck's constant, $\kb$ is Boltzmann's constant, $c$ is the light velocity and $T$ is the temperature. Using the framework of fluctuational electrodynamics~\cite{Rytovbook}, Polder and van Hove (PvH) were able to derive a general form of the RHT accounting for the optical properties of the media~\cite{Polder}. Since this seminal contribution, several reports have been published in the literature~\cite{Rytov, Loomis, Pendry, Volokitin2001, ZhangJAP,Bimonte2009}. A quantum-mechanical derivation~\cite{Holthaus} has confirmed these results obtained within the framework of fluctuational electrodynamics.  
While the first papers considered metals, it has been realized that the RHT at the nanoscale can be further enhanced for dielectrics due to the contribution of surface phonon polaritons~\cite{MuletEtAl2002,MuletAPL}. Recent reviews can be found in Refs.~\cite{Volokitin2007, SurfaceScienceReports,Dorofeyev3,ZhangReview}. 

The first attempts to measure a heat flux between metallic surfaces at room temperature and micrometric distances have proved to be inconclusive~\cite{XuJAP,Xu2}. Experiments in the nanometric regime have clearly demonstrated the transfer enhancement~\cite{Kittel,Wischnath}. Yet the lack of good control of the tip geometry did not allow quantitative comparison with theory. More recent experiments~\cite{NanolettArvind,NatureEmmanuel} are performed using silica taking advantage of the flux enhancement due to the resonant contribution of surface phonon polaritons. A good agreement between PvH theory and experiments  has been reported~\cite{NatureEmmanuel}. 

The purpose of this paper is to establish a link between the PvH form of the radiative heat flux and the formalism of transport in mesoscopic physics.  It will help to develop a more physical understanding of the RHT at the nanoscale, which also clarifies how losses and non-local effects determine the maximal achievable heat flux~\cite{ZhangJAP}. 
Finally, we will show that this reformulation raises the prospect of observing quantized conductance for systems with sizes on the order of the thermal wavelength  $\lambda_{\rm th}$.

\begin{figure}[Hhbt]
  \epsfig{file=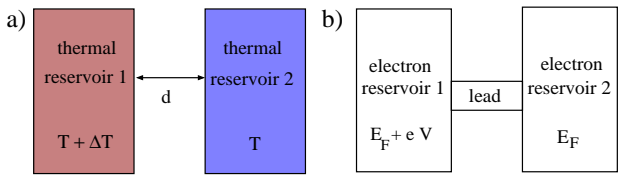, width=0.45\textwidth}
  \caption{\label{Fig:Landauer} Sketch of the conductance geometry:  a) Two thermal reservoirs are separated by a vacuum gap with width $d$, b) two electron reservoirs with a voltage difference V are connected through a nanowire.}
\end{figure}

We start our discussion with the PvH form of the RHT. We consider a vacuum gap with width $d$ separating two homogeneous half spaces labeled medium 1 and 2 [see Fig.~\ref{Fig:Landauer} a)]. Then, the heat flux is\cite{Polder,SurfaceScienceReports}
\begin{equation}
\begin{split}
\Phi(d,T_1,T_2) &= \int_{0}^{\infty} \frac{d \omega}{2\pi} [\Theta(\omega,T_1)-\Theta(\omega,T_2)] \\
                &\quad\sum_{j=s,p} \biggl[
                                    \int_{0}^{k_0} \frac{ \rd^2 \boldsymbol{\kappa}}{4\pi^2} \frac{(1-|r^{1}_j|^2)(1-|r^{2}_j|^2)}{|1 - r^1_j r^2_j e^{-2 i\gamma d}|^{2}} \\
                &\quad  + \int_{k_0}^{\infty} \frac{\rd^2 \boldsymbol{\kappa}}{4\pi^2} \frac{4 \Im(r^{1}_j)\Im(r^2_j)e^{-2\Im(\gamma) d} }{|1-r^1_j r^2_j e^{-2 \Im(\gamma) d}|^{2}}  \biggr]
\label{fluxtotal}
\end{split}
\end{equation}
where $\boldsymbol{\kappa} =\! (k_x, k_y)$ and $\gamma =\! \sqrt{k_0^2 - \kappa^2}$  are the parallel and normal wave vector, $\Theta(\omega,T)~=~\hbar\omega/[\exp(\hbar\omega/\kb T)~-~1)]$ is the mean energy of a harmonic oscillator, $k_0=\omega/c$, $r_j^{1,2}$ are the usual Fresnel factors for $s$- or $p$-polarized waves and the sum is carried out over $j=s~(TE) ,p~(TM)$ and accounts for the two polarizations. 
In this expression, we have clearly separated the propagating ($\kappa<k_0$) from the evanescent wave contribution ($\kappa>k_0$). The first one can be shown to be described by the usual radiometric approach~\cite{SurfaceScienceReports}. In particular, the term $1-|r_j^{1}|^2$ which is the interface transmittivity is exactly the emissivity of the interface of medium 1. It is thus tempting to extend the definition of the emissivity to evanescent waves. 
This has been proposed in Ref.~\cite{MuletEtAl2002} where $\Im(r_j^1)$ was introduced as a generalized emissivity. 
Obviously, the generalized emissivity is larger than $1$ since it accounts for an enhanced RHT. A drawback of this approach is that it does not provide any physical insight in the meaning of an emissivity larger than 1.

A different interpretation regarding the evanescent contribution only, was given by Pendry~\cite{Pendry} using 
the single interface local density of states (DOS) $\Im(r_j^1)\exp(- 2 \Im(\gamma) z)|_{z = 0}$ in the space ($\boldsymbol{\kappa}, z, \omega$). 
Here, the term local refers to the fact that we introduce the DOS in direct space at a distance z above the interface. 
To summarize, the first approach suggests to consider that   $\Im(r_j^{1,2})$ 
is a generalized emissivity whereas the second approach emphasizes the fact that  $\Im(r_j^{1,2})$ is proportional to the local DOS.
 
The interpretation in terms of local DOS is interesting as it indicates that the enhancement of the RHT at the nanoscale is due to an enhancement 
of the DOS.
Hence, it would be enlighting to develop a formulation of the RHT that highlights the number of modes (NOM) involved. Such a formulation has been developed in the context of charge transport in mesoscopic physics and is known as Sharvin conductance when the contact area is much larger than the electron wavelengths and Landauer formalism when the size is on the order of the wavelength so that the NOM involved in the transport becomes discrete.

Before, reformulating the heat flux in a Landauer-like way, let us remind the structure of the electrical conductivity in the mesoscopic regime. When considering that the dephasing length and the mean free paths are much larger than all the relevant length scales, the electrons can be described by their wave functions. Hence, the conduction through a constriction or nanowire [see Fig.~1 b)] can be  described by a scattering matrix that connects the incoming channels to the outgoing channels. It follows that the charge transport can be described by means of a sum over all the modes~\cite{Datta1995,Agrait}
\begin{equation}
  I = \frac{2 e}{h} \int \!\! \rd E\, \sum_nT_n(E) [f_1(E) - f_2 (E)]
\end{equation} 
where $e$ is the electron charge,  $T_n(E)$ is
the transmission probability of a mode $n$ and $f_1(E)=1/[\exp[(E - E_\rF - eV)/\kb T]+1]$ and $f_2(E)=1/[\exp[(E - E_\rF)/\kb T]+1]$ are the Fermi-Dirac distributions and $E_F$ the Fermi energy.
 For small applied voltages $V$, we obtain
\begin{equation}
  I = \frac{2 e^2}{h} \int \!\! \rd E\, \sum_n T_n(E) \biggl(-\frac{\partial f_0(E)}{\partial E} \biggr) V
\label{Eq:LandauerLinearResponse}
\end{equation}
where $2e^2/h$ is the quantum of conductance and $f_0 = f_2$. For temperatures $\kb T \ll E_\rF$, this result reduces to the Landauer formula
\begin{equation}
 I = \frac{2 e^2}{h} \sum_nT_n(E_\rF) V.
\label{Eq:Landauer}
\end{equation}

To derive such a formulation for the RHT case, we consider a situation where the two temperatures are close enough so that $T_1 = T + \Delta T$
and $T_2 = T$   
assuming that $\Delta T \ll T$. Then, we can write the flux $\Phi$ as
\begin{equation}
  \Phi = \sum_{j = s,p} \int\!\frac{\rd \omega}{2 \pi}\, \biggl[ \frac{\partial}{\partial T} \Theta(\omega,T) \biggr] \Delta T 
         \int \!\frac{\rd^2 \boldsymbol{\kappa}}{4 \pi^2}\, T^{12}_j (\omega,\kappa,d),
\label{Eq:flux}
\end{equation}
 Here, we have introduced the notation
\begin{equation}
   T^{12}_j (\omega,\kappa,d) = \begin{cases}  
                                  \frac{(1 - |r_j^1|^2)(1 - |r_j^2|^2)}{|D_j|^2} &, \kappa \leq k_0 \\
                                  \frac{4 \Im(r_j^1)\Im(r_j^2) \re^{- 2 \Im(\gamma) d}}{|D_j|^2} &, \kappa > k_0 
                                \end{cases} ,
\label{Eq:DefI0b}
\end{equation}
where $D_j = 1 - r_j^1 r_j^2 \exp(2 \ri \gamma d)$
is the Fabry-P\'{e}rot type denominator. For propagating modes, $T^{12}_j (\omega,\kappa,d)$ is the two interfaces transmission factor (TF) and takes values in the interval $[0,1]$. It follows that the heat flux is limited by the
black-body result. Hence, the expression strongly suggests to consider that  $T^{12}_j (\omega,\kappa,d)$ is also a TF for the evanescent modes. As shown in Ref.~\cite{Pendry} $T^{12}_j$ is indeed smaller than 1. This maximal value is reached if $|r_j^1||r_j^2|=e^{2 \gamma d}$ which for example can be satisfied 
for s- and p-polarized modes in the case of frustrated total internal reflection~\cite{jqsrt} when considering dielectrics and also for coupled surface modes (CSM).


By comparing the expression in Eq.~(\ref{Eq:LandauerLinearResponse}) with Eq.~(\ref{Eq:flux}) 
one can see that we have already formulated the heat flux equation in a very similar 
manner. The sum over transverse modes is given by the integral $\int \rd^2 \boldsymbol{\kappa}/4\pi^2$.
In order to get a manifestly Landauer-like structure, we need to integrate the modes over all energies. 
We first introduce the dimensionless variable $u = \hbar \omega/\kb T$. 
Then, by interchanging the order of integration we cast the heat flux in the form
\begin{equation}
  \Phi = \frac{\pi^2}{3} \frac{\kb^2 T}{h} \biggl( \sum_{j = s,p} \int\frac{\rd^2  \boldsymbol{\kappa}}{(2 \pi)^2} \overline{T}^{12}_j \biggr) \Delta T
\label{Eq:LandauerHeat}
\end{equation} 
where $\pi^2 \kb^2 T/3h$ is the universal quantum of thermal conductance~\cite{Pendry1983,RegoEtAl1999}. Here, we have introduced the mean transmission factor (MTF)
\begin{equation}
\label{transmission}
  \overline{T}^{12}_j =\frac{ \int_0^\infty \!\! \rd u\, f(u)\; T^{12}_j (u,\kappa,d)}{\int_0^{\infty}\!\! \rd u \, f(u)},
\end{equation}
where $f(u)=u^2 e^u/(e^u-1)^2$ and $\int_0^{\infty} f(u)\rd u=\pi^2/3$. This is a new quantity which resembles the transmission probability for the electrons in the
Landauer formula in Eq. (\ref{Eq:Landauer}). When dealing with electrons, only the TF at the Fermi energy is relevant because of the particular form of the Fermi-Dirac distribution. By contrast, when dealing with bosons, we need to introduce a TF averaged over all energies. 
\begin{figure}[ht]
  \epsfig{file=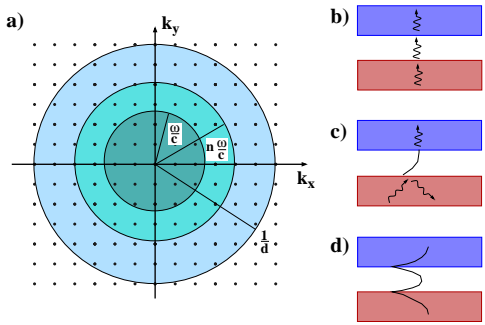, width=0.35\textwidth}
  \caption{\label{fig2}   Illustration of the contribution of b) propagating modes ($\kappa<k_0$), c) modes propagating in the dielectric but evanescent in the gap ($k_0<\kappa<nk_0$) and d) evanescent modes confined in the gap ($\kappa>nk_0$).   }
\end{figure}
Let us stress that $\overline{T}^{12}_j$ is always smaller than $1$. This property follows directly from the same property of $T^{12}_j$
and the definition of the MTF. Hence, we have now a new interpretation of the physical meaning of the term involving $\Im(r_j^{1,2})$ in Eq.~(\ref{fluxtotal}). 
Instead of interpreting $\Im(r^{1,2}_j)$ as the local DOS or generalized emissivity of a single interface, we consider now the two interface system (the gap) and define a TF averaged over all energies for a mode with specified $ \boldsymbol{\kappa}$. In this picture, the enhancement of the heat flux appears as the consequence of the increase of the NOM contributing to the RHT. We provide in Fig.~\ref{fig2} a) a schematic representation of the modes in the ($k_x, k_y$) plane, from which it becomes clear that the number of transverse modes diverges. For $\kappa<k_0$, we have the usual thermal radiation due to propagating modes, for $k_0<\kappa<nk_0$, there is a contribution of modes which can be viewed as frustrated total internal reflection when dealing with dielectrics. Finally, for $\kappa>nk_0$, we have the contribution of CSM confined in the gap. It remains to be studied under which conditions the TF takes significant values.


\begin{figure}[ht]
    \vspace*{-0.2cm} 
       \epsfig{file=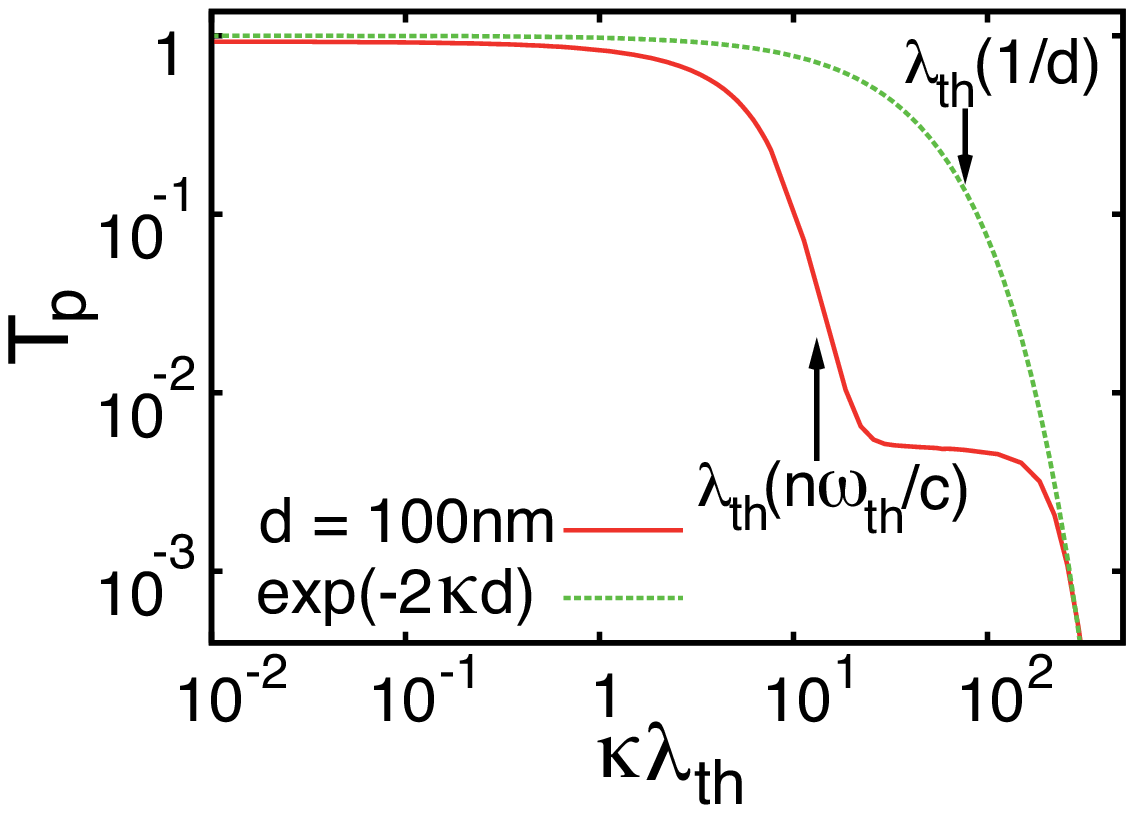, width=0.33\textwidth}
    \caption{\label{Fig:TransmissionFactor} 
      Plot of $\overline{T}^{12}_\rp$ and $\exp(-2 \kappa d)$ for $100\,{\rm nm}$
      choosing temperature $T = 300\,{\rm K}$ and two SiC~\cite{ShchegrovEtAl2000} slabs.
      The modes with $\kappa<nk_0$ have $\overline{T}^{12}_\rp \approx 1$. The surface phonon polaritons 
      with $\kappa > nk_0$ have a relatively small MTF but give the dominant contribution to the heat flux
      due to the large number of contributing modes. The MTF shows for very large $\kappa$ an exponential cutoff 
      $\propto \exp(-2 \kappa d)$.  
  }
  \vspace*{0.3cm}

   \epsfig{file=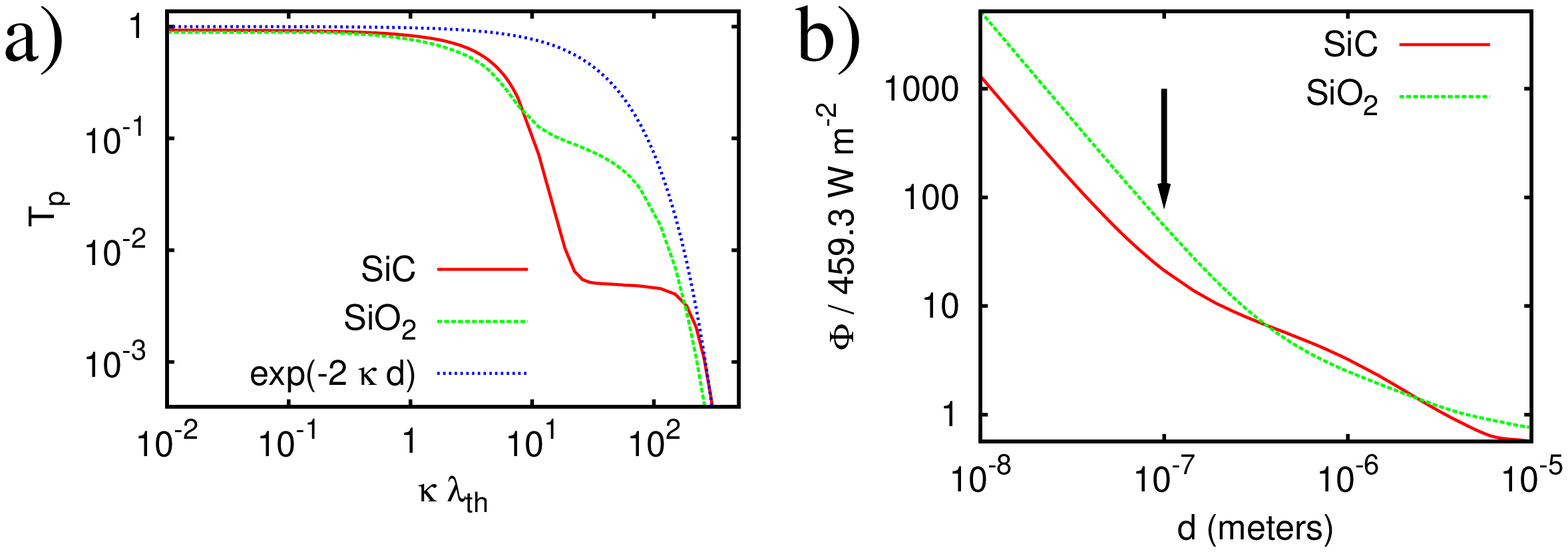, width=0.5\textwidth}
  \caption{\label{Fig:TransmissionFactorMat} Plot of a) $\overline{T}^{12}_\rp$ ($d = 100\,{\rm nm}$
     and $T = 300\,{\rm K}$) and b) resulting heat flux $\Phi$ for SiC and silica~\cite{Palik1985} normalized to the black body value $\Phi_{\rm BB} = 459.3\,{\rm W}{\rm m}^2$.
  }
\end{figure}

A detailed study of the TF $T(\omega, \kappa,d)$ is reported in~\cite{BiehsRousseauGreffetsuppl2010}.  
The key results are as follows. For a gap thickness $d$ larger  
than the thermal wavelength $\lambda_\rth$, the TF is negligible for  
evanescent waves ($k_0 < \kappa$) and oscillates between $0$ and $1$ with  
frequency for propagating waves. It is a Fabry-P\'{e}rot type behaviour.  
For a gap thickness of $100\,{\rm nm}$,  the  TF tends 
to $1$ for both propagating waves and waves with frustrated total reflection  
($k_0 < \kappa < n k_0$) thereby contributing to a significant enhancement of  
the flux. As the TF tends indeed to one for all  
frequencies (for s- and p-polarization), the contribution of these modes is simply given by the  
DOS $\approx 2\cdot(n^2 \lambda_\rth^{-2}/4 \pi)$ times the thermal quantum of conductance in  
agreement with previously reported results \cite{ZhangDoping,jqsrt}.

The value of the TF $T(\omega, \kappa,d)$ for CSM  
has a more subtle structure. It is essentially negligible  
everywhere in the ($\omega$,$\kappa$) plane except along the surface mode  
dispersion relation where it is close to $1$ (see Fig.~$1$ in~\cite{BiehsRousseauGreffetsuppl2010}). Hence, after averaging   
over frequencies, the MTF drops by two orders of magnitude as seen 
in Fig.~\ref{Fig:TransmissionFactor}.
We also see in Fig.~\ref{Fig:TransmissionFactorMat}~a)  
that this drop is smaller for silica than for SiC. This is because in  
the case of silica, there are more different surface modes contributing  
at different frequencies. 
Despite the low value of the MTF for surface modes, the number of additional 
surface modes is so large that the RHT is dominated by the CSM  
contribution at small distance as seen in Fig.~\ref{Fig:TransmissionFactorMat}~b). 

It is clear from the above discussion that the enhanced flux is due to  
the contribution of surface modes with a large value of $\kappa$. 
In~\cite{BiehsRousseauGreffetsuppl2010} we discuss the cutoff value of $\kappa$ 
which limits the contribution of the CSM in detail. 
It is roughly given by $1/d$ as indicated in the graph of Fig.~\ref{Fig:TransmissionFactor}. 
A more precise discussion  
shows~\cite{BiehsRousseauGreffetsuppl2010} that the losses play a key role in defining the exact position  
of the cutoff. It is given for $T(\omega, \kappa,d)$ at the surface resonance 
by $\kappa > \log[2/\Im(\epsilon)]/d$ which also sets a cutoff for the MTF. 

We finally turn to the ultimate limit  of the RHT. So far,  
it seems that the RHT conductance can diverge as $1/d^2$, since
the DOS $\propto 1/d^2$.  
Yet, it is known that there is a cutoff value for the spatial  
wave vectors of the phonons given by $\pi/a$ where $a$ is the lattice  
constant. This is equivalent to accounting for non-local properties of  
the material~\cite{ChapuisEtAl2008,JoulainHenkel2006}.

\begin{figure}[Hhbt]
  \epsfig{file=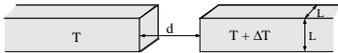, width=0.25\textwidth}
  \caption{\label{Fig:experiment} Sketch of a setup suitable for measuring the quantized radiative conductance between two wires with a cross section $L\times L$.}
\end{figure}

We now discuss an immediate consequence of our formulation. If we consider a system as depicted in Fig.~\ref{Fig:experiment}, we observe that the two wires have a finite transverse cross section $L \times L$. Hence, the electromagnetic modes are quantized for $L$ on the order of $\lambda_{\rm th}$ as shown for surface plasmon polaritons in Ref.~\cite{Yannick} and one expects to observe a radiative conductance of the gap with a discrete NOM similar to the quantized conductance of an aperture~\cite{Saenz,Montie}. An experiment similar to the work reported in Refs.~\cite{SchwabEtAl2000,MeschkeEtAl2006} should be feasible. 

%

%
To summarize, we have presented a Landauer-like reformulation for the nanoscale RHT 
putting it on the same footing as the mesoscopic electron transport. In particular, we have
introduced a mean transmission coefficient which is shown to fullfill all the properties needed to
give a clear understanding of the heat flux on nanoscale in terms of the DOS or NOM. 
In addition, we have proposed a near-field heat transfer experiment for measuring the quantized conductance.

%
%

\begin{acknowledgments}
S.-A.\ B. gratefully acknowledges support from the Deutsche Akademie der Naturforscher Leopoldina
(Grant No.\ LPDS 2009-7). J-J.\ G.\ and E.\ R.\ acknowledge the support of \textit{Agence Nationale de la Recherche} through \textit{Monaco} projects and Leti-Carnot Institute.
\end{acknowledgments}

%
%

\appendix

%
%

%
%

\end{document}